% ------------------------------------------------------------------
%    cd.tex   11 Feb 1999 // 9 April 1999 // 3 June 1999   
% ------------------------------------------------------------------
\def\ptitle{Spectral comparison theorem for the Dirac equation}
% ------------------------------------------------------------------------
% Change list publication -> preprint
% (1) Magnification : either magstep1 and 10pt fonts, or no mag and 12pt
% (2) Remove \parskip and set \baselineskip = 16pt before Abstract
% (3) Remove \np before (but not after) references
% (4) Add \input psfig.sty (or some such) at top
%     Add \psfig commands for each figure
% ------------------------------------------------------------------------
\nopagenumbers
%\magnification=\magstep1
\hsize 6.0 true in 
\hoffset 0.25 true in 
% 6 in width with 1.25 in margins default = (6.5, 0)
\emergencystretch=0.6 in                 % TEXBook p 107 : allows h-space 
\vfuzz 0.4 in                            % page-length flexibility
\hfuzz  0.4 in                           % line-length flexibility
\vglue 0.1true in
\mathsurround=2pt                        % Default is 2pt
\topskip=24pt                            % Default is 10pt
\def\nl{\noindent}                       % New line after equations
\def\np{\hfil\vfil\break}                % New page
\def\title#1{\bigskip\noindent\bf #1 ~ \tr\smallskip} % Headings
% --------------------------------------------------------------------
%   PC Fonts (used only locally)
% --------------------------------------------------------------------
%\font\tr=TIMENRR                       % Times New Roman: see output
%\font\bf=TIMENRB                       % Redefinition
%\font\it=TIMENRRI                      % Redefinition       
%\font\trbig=TIMENRR scaled \magstep3   % For main Title
%\font\th=CMBXSL10                      % Theorems                    
%\font\tiny=CMBX8                       % Page numbers
% --------------------------------------------------------------------
%  generic unix fonts (lower case names)
% --------------------------------------------------------------------
\font\tr=cmr12                          % Our default
\font\bf=cmbx12                         % Redefinition
\font\sl=cmsl12                         % Redefinition
\font\it=cmti12                         % Redefinition
\font\trbig=cmbx12 scaled 1500          % Main Title
                          % Theorems                       
\font\tiny=cmr10                         % Running title
% --------------------------------------------------------------------
             % Math Sets eg R -> |R
                 % bold in math mode
                % small bold in math mode
\def\ng{>\kern -9pt|\kern 9pt}          % not greater than
                       % bra  <  math mode
                       % ket  >  math mode
\def\hi#1#2{$#1$\kern -2pt-#2}          % hyphen \hi{N}{body} = N-body
\def\hy#1#2{#1-\kern -2pt$#2$}          % hyphen hy{large}{N} = large-N

\def\half{{1 \over 2}}
\def\kr{{{\tau k} \over {r}}}
% ----------------------------------------------------------------------

  %  QED
 % SQUARE 
% ---------------------------------------------------------------------- 
\output={\shipout\vbox{\makeheadline
                                      \ifnum\the\pageno>1 {\hrule}  \fi 
                                      {\pagebody}   
                                      \makefootline}
                   \advancepageno}

\headline{\noindent {\ifnum\the\pageno>1 
                                   {\tiny \ptitle\hfil page~\the\pageno}\fi}}
\footline{}
% ---------------------------------------------------------------------
\newcount\zz  \zz=0  % switch for printing references
\newcount\q   %  reference number
\newcount\qq    \qq=0  % starting reference number-1   (usually zero)

\def\pref #1#2#3#4#5{\frenchspacing \global \advance \q by 1     % paper reference
    \edef#1{\the\q}
       {\ifnum \zz=1 { %
         \item{[\the\q]} 
         {#2} {\bf #3},{ #4.}{~#5}\medskip} \fi}}

\def\bref #1#2#3#4#5{\frenchspacing \global \advance \q by 1     % book reference
    \edef#1{\the\q}
    {\ifnum \zz=1 { %
       \item{[\the\q]} 
       {#2}, {\it #3} {(#4).}{~#5}\medskip} \fi}}

\def\gref #1#2{\frenchspacing \global \advance \q by 1  % general reference
    \edef#1{\the\q}
    {\ifnum \zz=1 { %
       \item{[\the\q]} 
       {#2}\medskip} \fi}}

 \def\sref #1{~[#1]}

\def\references#1{\zz=#1
   \parskip=2pt plus 1pt   % default is 0pt plus 1pt       
   {\ifnum \zz=1 {\noindent \bf References \medskip} \fi} \q=\qq

\bref{\reed}{M. Reed and B. Simon}{Methods of Modern Mathematical Physics IV: Analysis of Operators}{Academic, New York, 1978}{The min-max principle for the discrete spectrum is discussed on p75}
  \bref{\thir}{W. Thirring}{A Course in Mathematical Physics 3: Quantum Mechanics of Atoms and Molecules}{Springer, New York, 1981}{The min-max principle for the discrete spectrum is discussed on p152}
\pref{\rosen}{M. E. Rose and R. R. Newton, Phys. Rev.}{82}{470 (1951)}{}
\bref{\mes}{A. Messiah}{Quantum Mechanics vol II}{North-Holland, Amsterdam, 1961}{The Dirac equation for the Coulomb problem is discussed on p930}
\bref{\ros}{M. E. Rose}{Relativistic Electron Theory}{J. Wiley, New York, 1961}{The Dirac equation for central fields is discussed on p159; the occurrence of nodes in the wave functions for potentials that are Coulombic near $r = 0$ is discussed on p169.}
\pref{\hella}{H. Hellmann, Acta. Physiochem. URSS I}{6}{913 (1935)}{}
\pref{\hellb}{H. Hellmann, Acta. Physiochem. URSS IV}{2}{225 (1936)}{}
\bref{\hellb}{H. Hellmann}{Einf\"urung in die Quantenchemie}{F. Denticke, Leipzig, 1937}{p. 286}
\pref{\fyn}{R. P. Feynman, Phys. Rev.}{56}{340 (1939)}{}
\pref{\meht}{C. H. Mehta and S. H. Patil, Phys. Rev. A}{17}{34 (1978)}{}
\pref{\halla}{R. L. Hall, Phys. Rev. A}{32}{14 (1985)}{}  % NR Screened Coulomb
\pref{\hallb}{R. L. Hall, J. Phys. A.}{19}{2079 (1986)}{} % R  Screened Coulomb
\pref{\piep}{W. Pieper and W. Greiner, ~Z. Physik}{218}{327 (1969)}{} % sqr well 
\bref{\grein}{W. Greiner}{Relativistic Quantum Mechanics: Wave Equations}{Springer-Verlag, Heidelberg, 1990}{Solutions of the Dirac equation for the finite square well in three dimensions are discussed on p. 174; the non-monotone dependence of the energy on coupling to a {\it scalar} potential in one dimension is shown on p. 168.}

 }

 \references{0}    % Initialization of reference numbers
% ------------------------------------------------------------ end our ref.tex ---------------

% ------------------------------------------------------ 
%   Title page and Abstract
% ------------------------------------------------------
\tr % our standard font
% \rightline{To appear in \bf PRL}
\vskip 1.0true in
\centerline{\trbig Spectral Comparison Theorem}
\vskip 0.3true in
\centerline{\trbig for the Dirac Equation}
\vskip 0.5true in
\baselineskip 12 true pt % for address only                                     
\centerline{\bf Richard L. Hall}\medskip
\centerline{\sl Department of Mathematics and Statistics,}
\centerline{\sl Concordia University,}
\centerline{\sl 1455 de Maisonneuve Boulevard West,}
\centerline{\sl Montr\'eal, Qu\'ebec, Canada H3G 1M8.}
\vskip 0.2 true in
\centerline{email:\sl~~rhall@cicma.concordia.ca}
\bigskip\bigskip

% ---------------------------------------------------------------------
% spacing: remove next line for preprint version
%\parskip=5pt plus 1pt      % MAIN PARSKIP remove for preprint
\baselineskip = 18true pt  % baselineskip paper 18 preprint 16
% ---------------------------------------------------------------------
\centerline{\bf Abstract}\medskip
% ---------------------------------------------------------------------
We consider a single particle which is bound by a central potential and obeys the Dirac equation.   We compare two cases in which the masses are the same but $V_a < V_b,$ where $V$ is the time-component of a vector potential.  We prove generally that for each discrete eigenvalue $E$ whose corresponding (large and small) radial wave functions have no nodes, it necessarily follows that $E_a < E_b.$  As an illustration, this general relativistic comparison theorem is applied to approximate the Dirac spectrum generated by a screened-Coulomb potential.

\medskip\noindent PACS~~03.65.Ge,~03.65.Pm,~31.15.Bs,~02.30.Mv.

\np
% ------------------------------------------------------ 
  \title{1.~~Introduction}
% ------------------------------------------------------
The comparison theorem of non-relativistic quantum mechanics states that

$$V_a < V_b \ \Rightarrow \ E_a < E_b,\eqno{(1.1)}$$

\nl where $V$ is an attractive potential which supports discrete eigenvalues.  This theorem is usually proven for every eigenvalue by an application of the variational (min-max) characterization\sref{\reed, \thir} of the discrete part of the Schr\"odinger spectrum. Such a proof is unavailable for the corresponding Dirac problem since the Dirac Hamiltonian is not bounded below, and the spectrum cannot be defined variationally. An early and very detailed analysis of the Dirac spectrum for central potentials has been given by Rose and Newton\sref{\rosen}. The impossibility of a general proof of a comparison theorem for the Dirac problem has led to the commonly-held belief that no such theorem at all could be established.   In this paper we prove that for attractive central potentials, (1.1) is valid for each discrete Dirac eigenvalue whose wave functions have no nodes, that is to say, for the bottom of each angular-momentum subspace.   

The possibility of a theorem of this kind was suggested by an attempt to prove a comparison theorem for the non-relativistic problem {\it without} the use of min-max. If we write the two Schr\"odinger comparison Hamiltonians in dimensionless form as $H_a = -\Delta + V_a$ and $H_b = -\Delta + V_b$ and we write down the corresponding eigen equations in one dimension we get:

$$-\psi''(x) + V_a(x)\psi(x) = E_a \psi(x),\eqno{(1.2)}$$
\nl and
$$-\phi''(x) + V_b(x)\phi(x) = E_b \phi(x).\eqno{(1.3)}$$

\nl We now assume that the wave functions are normalized, we form the difference (1.2)$\phi$~-~(1.3)$\psi,$ and integrate it over all space to obtain:

$$\int_{-\infty}^{\infty}(V_a(x) - V_b(x))\psi(x)\phi(x)dx = (E_a - E_b)\int_{-\infty}^{\infty}\psi(x)\phi(x)dx.\eqno{(1.4)}$$

\nl This equation immediately establishes (1.1), provided the wave functions have no nodes.  If the potentials are symmetric, and we consider the lowest odd state, then the wave functions vanish at the origin and we obtain (1.4), with integrations on $[0, \infty)$; and this again proves (1.1).  Thus the theorem is established also for the bottom of the odd-parity space.  Consequently this result extends to the corresponding radial problem in $N > 1$ dimensions, provided that we consider nodeless states at the bottom of each angular-momentum subspace.   

In Section~(2) we apply similar reasoning to the Dirac problem and we are able to prove that the comparison theorem (1.1) is valid for every discrete eigenvalue which is at the bottom of an angular-momentum subspace.  In Section~(3) the new relativistic comparison theorem is applied to approximate the spectrum produced by a screened-Coulomb potential: a set of soluble comparison potentials are generated by use of the `potential envelope method'.  
% ------------------------------------------------------
  \title{2.~~The relativistic comparison theorem}
% ------------------------------------------------------
We consider a single Dirac particle moving in a central vector potential with time component $V(r),$ and a fixed mass $m.$  Since we are not able to accommodate variations in the scalar potential in the comparison theory presented here we do not allow for it at this stage; a more detailed remark will be made later concerning this question.  We adopt a notation similar to that of Messiah\sref{\mes} and Rose\sref{\ros}.  We let $\psi_1$ and $\psi_2$ be the `large and small' radial wave functions used to construct the Dirac spinor corresponding to a total angular momentum of $j.$  We employ the variables $\tau = \pm 1,$ and $k = j + \half,$ so that the parity $P$ of an energy eigenstate is given by $P = (-1)^{j+{\tau \over 2}} = \pm.$  If the eigenvalues are labelled $E^{P}_{n j},$ where $n = 1,2,3,\dots,$ enumerates the distinct radial states, then the degeneracy of this energy symbol is exactly $2j+1.$  In this notation the principal quantum number for the Coulomb problem becomes $\nu = n + k - \half(1-\tau).$  Meanwhile, the boundary conditions and normalization we have adopted for the radial functions are:

$$\psi_1(0) = \psi_2(0) = 0, \quad \int_{0}^{\infty}(\psi_{1}^{2}(r) + \psi_{2}^{2}(r))dr = 1.\eqno{(2.1)}$$

We now consider two different potentials $V_a$ and $V_b$ and we associate with them the corresponding pairs of radial functions $\{\psi_1, \psi_2\}$ and $\{\phi_1, \phi_2\}$ and we obtain a pair of coupled radial equations\sref{\ros} for each problem:

$$\psi_{2}' - \kr\psi_2 = (m + V_a - E_a)\psi_1\eqno{(2.2)}$$

$$\psi_{1}' + \kr\psi_1 = (m - V_a + E_a)\psi_2\eqno{(2.3)}$$

$$\phi_{2}' - \kr\phi_2 = (m + V_b - E_b)\phi_1\eqno{(2.4)}$$

$$\phi_{1}' + \kr\phi_1 = (m - V_b + E_b)\phi_2.\eqno{(2.5)}$$

\nl We consider now the special combination of these four equations which is given by multiplying each equation by a wave function and summing, according the the prescription

$$(2.2)\phi_1 + (2.5)\psi_2 - (2.3)\phi_2 - (2.4)\psi_1,$$

\nl and we find:

$$(\phi_1\psi_2)' - (\psi_1\phi_2)' = (\phi_1\psi_2 + \psi_1\phi_2)[V_a - V_b - (E_a - E_b)].\eqno{(2.6)}$$

\nl By integrating (2.6) and using the boundary conditions we obtain:

$$\int_{0}^{\infty}(\phi_1\psi_2 + \psi_1\phi_2)[V_a - V_b]dr = [E_a - E_b]\int_{0}^{\infty}(\phi_1\psi_2 + \psi_1\phi_2)dr.\eqno{(2.7)}$$

\nl If the wave functions have no nodes, the factors involving them on each side of (2.7) have the same sign: hence, under these conditions, this equation establishes the comparison theorem (1.1) for the Dirac problem. It should perhaps be mentioned here that in order to derive the comparison result from (2.7), it is necessary to assume that the potentials and the eigenvalues are both real: if, for example, potential parameters stray into regions where a corresponding eigenvalue becomes complex, then (2.7) would no longer imply (1.1), since the complex numbers are not well ordered.   

We now turn to the corresponding problem for scalar potentials.  Let us suppose that the vector potentials are the same and that the masses are given respectively by $m_{a}(r)$ and $m_{b}(r).$ The same type of reasoning as we have used above leads to the expression:

$$\int_{0}^{\infty}(\phi_1\psi_1 - \phi_2\psi_2)[m_a - m_b]dr = [E_a - E_b]\int_{0}^{\infty}(\phi_1\psi_1 + \phi_2\psi_2)dr.\eqno{(2.8)}$$

It is clear that we can only draw the conclusion $E_a < E_b$ from (2.8) under inconvenient assumptions, such as dominance of the large radial component.  In the general case, even for node-free wave functions, it seems that no simple comparison theorem for scalar potentials can be derived in this way; indeed, Greiner\sref{\grein} has exhibited an example (in one dimension) in which the dependence of the energy on coupling to a scalar potential is not monotone.  In the non-relativistic limit, with constant masses, the well-known Hellmann-Feynman result\sref{\hella -\fyn} follows, as we would expect.
  
% ------------------------------------------------------
  \title{3.~~Energy upper bounds for a screened-Coulomb potential}
% ------------------------------------------------------
In order to study an application of the comparison theorem we need two potentials $V^{(t)}(r)$ and $V(r)$ which are ordered, say

$$V^{(t)}(r) \geq V(r).\eqno{(3.1)}$$

\nl We choose for $V(r)$ the screened-Coulomb potential suitable for large atoms which has been studied by Mehta and Patil\sref{\meht} and is given by

$$V(r) = -\left({v \over r}\right)[1 - r\lambda (1-1/Z)/(1+\lambda r)],\eqno{(3.2)}$$
\nl where
$$v = \alpha Z \quad {\rm and} \quad \lambda = 0.98\alpha Z^{1 \over 3}.\eqno{(3.3)}$$

\nl For the comparison potential $V^{(t)}(r)$ we generate not one, but a {\it set} of `tangential' potentials by using the method of `potential envelopes'\sref{\halla, \hallb}.  The apparatus of this theory is not essential to the illustration so long as (3.1) is valid. We now give a short self-contained derivation of this set of comparison potentials, and we also provide an independent verification of (3.1).     

The envelope method requires a soluble base potential which we take to be the pure Hydrogenic potential $-u/r = uh(r).$ This potential leads to a discrete spectrum which, in units of $mc^2$ and for $u < 1,$ is given exactly\sref{\ros} by

$$D^{P}_{n j}(u) = D(u) = \left\{1 + u^{2}\left[n - \half(1-\tau) + (k^{2} - u ^{2})^{\half}\right]^{-2}\right\}^{-\half},\eqno{(3.4)}$$

\nl where $k = j + \half,$ and $n = 1,2,3,\dots,$ counts the discrete eigenvalues for each given $\{\tau, j\}$ pair. 

If we write the potential (3.2) as the transformation $V(r) = g(h(r))$ of the pure Coulomb potential $h(r) = -1/r,$ then we have

$$g(h) = vh + v\lambda(1 - 1/Z)\left[1 + {\lambda \over {h - \lambda}}\right].\eqno{(3.5)}$$

\nl It follows immediately that $g'(h) > 0$ and $g''(h) < 0;$ that is to say, $g$ is monotone increasing and concave. As a consequence of this, every tangent line to $g(h)$ is a shifted Coulomb potential of the form

$$V^{(t)}(r) = A(t) + B(t)h(r) = \{g(h(t)) - h(t)g'(h(t))\} + g'(h(t))h(r),\eqno{(3.6)}$$

\nl where $r = t$ is the point of contact with $V.$  We now know that the potential inequality (3.1) is valid because the concavity of $g$ implies that it lies below its tangents.  For the present example we can also show, by a direct calculation, that the potential difference is given by the following clearly positive expression: 

$$V^{(t)}(r) - V(r) = {{v(1-1/Z)\lambda^{2}(r-t)^{2}}\over{r(1+\lambda r)(1+\lambda t)^2}} \geq 0.\eqno{(3.7)}$$

The eigenvalues corresponding to the shifted Coulomb potential (3.6) can be found exactly and are given immediately in terms of the known pure Hydrogenic eigenvalues $D(u)$ by

$${\cal E}^{(t)} = A(t) + D(B(t)) \geq E.\eqno{(3.8)}$$

\nl The inequality in (3.8) follows from the potential inequality (3.1) and our comparison theorem, provided the large and small radial functions are node free.  For potentials that are Coulombic near $r = 0,$ the argument of Rose\sref{\ros} demonstrates that the number of nodes is the {\it same} for $\psi_1$ and $\psi_2$ only if $\tau = -1.$  Hence we must restrict our considerations to the eigenvalues at the bottom of each angular-momentum subspace, that is to say, to those with $\tau = -1,\quad n = 1.$  All it remains to do is to minimize ${\cal E}^{(t)}$ with respect to $t > 0$ in order to obtain the best envelope approximation for each eigenvalue.  By a simple change of variable $t \rightarrow u = g'(h(t)),$ the best upper energy bound may be written in the much more compact form:

$$E^{U} = \min_{u \in (0,1)}\left\{D(u) - uD'(u) + V(-1/D'(u))\right\}.\eqno{(3.9)}$$  

The `principal quantum number' $\nu$ of the Coulomb problem may be defined generally by the expression $\nu = n + k - \half(1-\tau).$  Thus, for the states whose energies obey the comparison inequality, we have $\nu = k = j + \half,$ $P = (-1)^{k-1},$ and $\ell = j-\half,$ where $\ell$ is the orbital angular-momentum quantum number in the first two components of the Dirac spinor. The spectroscopic designation is then $\nu\ell_{j},$ where $\ell = \{0,1,2,\dots\} \sim  \{s, \ p, \ d,\dots\}$  In Table~(1) we exhibit some upper bounds $E^{U}$ found by Eq.(3.9), along with corresponding accurate approximations $E$ found numerically. 
\hfil\vfil\break % Kludge ******************** 
% ------------------------------------------------------
  \title{4.~~Conclusion}
% ------------------------------------------------------
A comparison theorem is a very useful general tool because it allows us to predict spectral ordering without actually having to solve the eigenvalue problems.  In the relativistic case we are restricted to the bottoms of the angular-momentum subspaces. Perhaps this limitation can be weakened in future.  Computations made with the screened-Coulomb potential have not revealed any counter example to the conjecture that (1.1) is generally true, for all the discrete eigenvalues.  Since the energy functions $D(u)$ for the Hydrogenic problem are monotone\sref{\hallb} in the coupling parameter $u,$ and the corresponding functions $E(V_{o})$ for the square well, studied by Pieper and Greiner\sref{\piep, \grein}, are monotone, a counter example is certainly not immediately available.  On the other hand, it is unlikely that a simple extension could be made to the proof given here so that it would apply also to states which have nodes, since this is not possible in the more regular Schr\"odinger case.  Further progress will probably have to await some kind of non-standard extension of min-max theory rich enough to accommodate the unbounded Dirac energy operator.  

As an illustration, we have shown that the new relativistic comparison theorem allows us to derive energy bounds such as (3.9). Formulas like this have the advantage that they describe approximately how the discrete spectrum depends on all of the potential parameters.  This quasi-analytical information is complementary to purely numerical calculations.     
% ------------------------------------------------------   
   \title{Acknowledgments}
% ------------------------------------------------------
Partial financial support of this work under Grant No. GP3438 from the Natural Sciences and Engineering Research Council of Canada is gratefully acknowledged. The author would also like to thank Jerrold Franklin for his comments on the presentation of the envelope method, Norman Dombey for pointing out the counter example of Greiner\sref{\grein} for scalar potentials, Walter Thirring and Jakob Yngvason for their comments on the comparison theorem, and Bernd Thaller for pointing out the paper by Rose and Newton\sref{\rosen}.  
\np
% ------------------------------------------------------ 
  \references{1}
% ------------------------------------------------------
\np
% ------------------------------------------------------
%  Table
% ------------------------------------------------------
\noindent {\bf Table 1}~~Upper bounds $E^{U}_j$ by the envelope method, and accurate numerical values $E_j,$ for the bottoms of the first two angular momentum subspaces labelled by $\tau = -1,$ and $j = \half$ and ${3\over 2}.$  The spectral descriptions of these two eigenvalues are respectively $1s_{\half}$ and $2p_{3\over 2}.$
\vskip 0.5 true in 

\def\vr{\vrule height 18 true pt depth 11 true pt}
\def\vra{\vr\hfill} \def\vrb{\hfill &\vra} \def\vrc{\hfill & \vr\cr\hrule}
\def\vrq{\vr\quad} 

$$\vbox{\offinterlineskip
 \hrule
\settabs
\+ \vrq \kern 0.4true in &\vrq \kern 0.7true in &\vrq \kern 0.7true in &\vrq \kern 0.7true in &\vrq \kern 0.7true in &\vr\cr\hrule

\+ \vra $Z$ \vrb $E^{U}_{\half}$\vrb $E_{\half}$\vrb $E^{U}_{3\over 2}$\vrb $E_{3\over 2}$\vrc
\+ \vra 20 \vrb -4.2571\vrb -4.3157\vrb\ -0.48522\vrb -0.53361\vrc 
\+ \vra 30 \vrb -10.2099\vrb -10.2960\vrb -1.3811\vrb -1.4659\vrc
\+ \vra 40 \vrb -18.9615\vrb -19.0732\vrb -2.8232\vrb -2.9448\vrc
\+ \vra 50 \vrb -30.7186\vrb -30.8543\vrb -4.8486\vrb -5.0070\vrc
\+ \vra 60 \vrb -45.7601\vrb -45.9189\vrb -7.4879\vrb -7.6825\vrc
\+ \vra 70 \vrb -64.4734\vrb -64.6545\vrb -10.7692\vrb -10.9997\vrc
\+ \vra 80 \vrb -87.4118\vrb -87.6148\vrb -14.7216\vrb -14.9877\vrc

}$$
% -------------------------------------------------------------------------
\hfil\vfil
\end